\documentclass[12pt,preprint]{aastex}
\shorttitle{SMA Observations of IRAM\,04191+1522}%
\shortauthors{X. Chen et al.}%

\begin{document}

\title{Discovery of A Binary System in IRAM\,04191+1522}

\author{Xuepeng~Chen\altaffilmark{1}, H\'{e}ctor~G.~Arce\altaffilmark{1}, Michael~M.~Dunham\altaffilmark{1}, and Qizhou Zhang\altaffilmark{2}}

\affil{$^1$Department of Astronomy, Yale University, Box 208101, New Haven, CT 06520-8101, USA; xuepeng.chen@yale.edu}%
\affil{$^2$Harvard-Smithsonian Center for Astrophysics, 60 Garden Street, Cambridge, MA 02138, USA}%

\begin{abstract}

We present high angular resolution observations of the Class\,0 protostar IRAM\,04191+1522, using 
the Submillimeter Array (SMA). The SMA 1.3\,mm continuum images reveal within IRAM\,04191+1522 
two distinct sources with an angular separation of 7.8\,$\pm$\,0.2$''$. The two continuum sources are 
located in the southeast-northwest direction, with total gas masses of $\sim$\,0.011\,$M_\odot$ and 
$\sim$\,0.005\,$M_\odot$, respectively. The southeastern source, associated with an infrared source 
seen in the $Spitzer$ images, is the well-known Class\,0 protostar with a bolometric luminosity of 
$\sim$\,0.08\,$L_\odot$. The newly-discovered northwestern continuum source is not visible in the 
$Spitzer$ images at wavelengths from 3.6 to 70\,$\mu$m, and has an extremely low bolometric 
luminosity ($<$\,0.03\,$L_\odot$). 
Complementary IRAM N$_2$H$^+$\,(1--0) data that probe the dense gas in the common envelope 
suggest that the two sources were formed through the rotational fragmentation of an elongated dense 
core. Furthermore, comparisons between IRAM\,04191+1522 and other protostars suggest that most 
cores with binary systems formed therein have ratios of rotational energy to gravitational energy 
$\beta_{\rm rot}$ $>$ 1\%. This is consistent with theoretical simulations and indicates that the level of 
rotational energy in a dense core plays an important role in the fragmentation process.

\end{abstract}


\section{INTRODUCTION}

The origin of binary stars is one of the major puzzles in our understanding of star formation.
Binary and higher-order multiple systems have been frequently observed in all stages of stellar 
evolution, but at present we do not understand well how this occurs (see, e.g., reviews by 
Tohline 2002; Goodwin et al. 2007).
The formation of binary systems begins in the earliest phase of the star formation process, 
when the young protostar, generally referred to as Class\,0 protostar (Andr{\'e} et al. 2000), is 
still deeply embedded within a dense core of gas and dust. In order to achieve a comprehensive 
understanding of binary star formation, high angular resolution observations of the gas and optically
thin dust emission at (sub-)\,millimeter wavelengths are therefore required to probe the system kinematics
and individual core masses.
However, these observations were long limited by the low angular resolution of single-dish millimeter 
telescopes, and have only become possible with the recent availability of large (sub-)\,millimeter 
interferometers.

In the past decade, a handful of Class\,0 protostellar binary (protobinary) systems have been found 
using different interferometers (e.g., Looney et al. 2000; Launhardt 2004), and there are increasing 
interferometric studies of binarity in the protostellar phase (e.g., Volgenau et al. 2006; Chen et al. 
2007; Maury et al. 2010). 
Nevertheless, due to the large effort and observing time required, the number of known and 
well-studied protobinary systems is still small. Furthermore, high angular resolution kinematics 
data, which could be used to study the physical process leading to the formation of binary stars, 
are also lacking. Therefore, it is critical to observe, at high angular resolution, more protostellar 
cores in nearby molecular clouds to search for protobinary candidates and study in detail their 
kinematic properties, and then compare the observational results with different theoretical models.

In this Letter, we present interferometric observations of IRAM\,04191+1522 (hereafter IRAM\,04191), 
using the Submillimeter Array\footnote{The Submillimeter Array is a joint project between the 
Smithsonian Astrophysical Observatory and the Academia Sinica Institute of Astronomy and Astrophysics 
and is funded by the Smithsonian Institution and the Academia Sinica.} (SMA; Ho et al. 2004).
IRAM\,04191 is a Class\,0 protostar discovered in the southern part of the Taurus molecular 
cloud (Andr{\'e} et al. 1999). From its low bolometric luminosity ($L_{\rm bol}$\,$\sim$\,0.15\,$L_\odot$) 
and temperature ($T_{\rm bol}$\,$\sim$\,18\,K), Andr{\'e} et al. (1999) estimated an age of 
$\sim$\,1--3\,$\times$\,10$^{4}$\,yr, suggesting IRAM\,04191 as one of the youngest protostars 
in Taurus. IRAM\,04191 drives a large scale ($\sim$\,0.1\,pc), bipolar molecular outflow, which 
extends in the northeast-southwest direction (Andr{\'e} et al. 1999; Lee et al. 2005). Various 
molecular line observations have shown subsonic infall, fast rotation, depletion, and deuteration 
in IRAM\,04191 (see, e.g., Belloche et al. 2002; Takakuwa et al. 2003; Belloche \& Andr\'{e} 2004; 
Lee et al. 2005). This source is also detected in the {\it Spitzer Space Telescope} ($Spitzer$) images 
at wavelengths between 3.6 and 70\,$\mu$m, and is classified as a very low luminosity object (VeLLO), 
due to its low internal luminosity\footnote{The internal luminosity is the luminosity of the central source, 
which excludes the luminosity arising from external heating.} ($L_{\rm int}$\,$\sim$\,0.08\,$L_\odot$; 
Dunham et al. 2006, 2008).
Based on the SMA observations, we report in this Letter that IRAM\,04191, regarded as an excellent 
example of isolated single star formation in the past decade, is actually a binary system, which may
in turn represent one of the best examples of a protobinary system formed though rotational fragmentation.

\section{OBSERVATIONS AND DATA REDUCTION}

The SMA 230\,GHz observations of IRAM\,04191 were carried out in 2007 November ($\sim$\,5 
hours integration time) with the compact configuration (PI: J.~Karr), and in 2011 March ($\sim$\,1 
hour integration time) with the subcompact configuration (PI: X.~Chen). Seven antennas were used 
in both configurations, providing baselines ranging from 9--53\,k$\lambda$ (compact) and 
6--51\,k$\lambda$ (subcompact), respectively.
In the 2007 observations, the digital correlator was set up to cover the frequencies from 219.5 to 
221.4\,GHz and from 229.5 to 231.4\,GHz in the lower and upper sidebands (LSB and USB), 
respectively. The 1.3\,mm continuum emission was recorded with a total bandwidth of $\sim$\,3.12\,GHz, 
combining the line-free portions of the two sidebands. In the 2011 observations, the SMA bandwidth 
was upgraded to 4\,GHz, and the correlator was set up to cover the frequencies from 216.8 to 220.8\,GHz 
(LSB) and from 228.8 to 232.8\,GHz (USB), respectively, resulting in a total continuum bandwidth of 
$\sim$\,7.46\,GHz. The SMA primary beam is about 55$''$ at 230\,GHz.

The visibility data were calibrated with the MIR package (Qi 2005). In the 2007 observations, the
quasar 3c454.3 was used for bandpass calibration, and quasars 3c120 and 0530+135 for 
gain calibration. In the 2011 observations, the quasar 0854+201 was used for bandpass calibration, 
and quasars 3c84 and 0359+509 for gain calibration. Uranus was used for absolute flux calibration
in the 2007 observations, while 0854+201 was used in the 2011 observations. We estimate
a flux accuracy of $\sim$\,20\% for the data, by comparing the final quasar fluxes with the SMA 
calibration database. The calibrated visibility data were further imaged using the Miriad toolbox
(Sault et al. 1995).

\section{RESULTS}

Figure~1 shows the SMA 1.3\,mm dust continuum images (top panel) and the $Spitzer$ infrared images 
(bottom panel) of IRAM\,04191. The SMA 1.3\,mm dust continuum observations reveal within IRAM\,04191 
two distinct sources, which are separated by 7.8\,$\pm$\,0.2$''$ in the northwest-southeast direction (see 
Figures~1a-c). These two sources were detected with the SMA compact and subcompact configurations 
independently, thus yielding a solid detection. The SMA compact configuration observations detected 
relatively compact emission clearly arising from two distinct sources (see Figure~1a), while the subcompact 
configuration observations detected relatively extended emission with two local maxima (see Figure~1b), 
likely tracing an inner common envelope.
As shown in the $Spitzer$ images (see Figures~1d-f), the southeastern continuum source is associated 
with an infrared source, and is therefore referred to as IRAM\,04191\,IRS in this work (see also Dunham 
et al. 2006). In contrast, the northwestern continuum source has no compact infrared emission at the 
$Spitzer$ bands from 3.6 to 70\,$\mu$m, and is referred to as IRAM\,04191\,MMS\footnote{The $Spitzer$
spatial resolution is $\sim$\,18$''$ at 70\,$\mu$m, which fails to resolve sources IRS and MMS. However,
due to the facts that (1) the 70\,$\mu$m emission peaks at source IRS, without any shift toward source
MMS, and (2) the emission is roughly circularly symmetric (see Figure~1f), we consider that all the 
70\,$\mu$m emission comes only from source IRS.}.
The comparisons with previous high angular resolution observations (e.g., Belloche et al. 2002; Lee et
al. 2005) suggest that source IRS is the well-known Class\,0 protostar, which drives a large scale bipolar 
outflow in the northeast-southwest direction.

The continuum fluxes of the two sources were derived from Gaussian fits to the restored images (using the
combined compact and subcompact data) with the Miriad command {\it imfit} (see Table~1). The fluxes of 
sources IRAM\,04191\,IRS and MMS are $\sim$\,9.3\,mJy and $\sim$\,4.7\,mJy, respectively. For comparison, 
the continuum flux detected in the IRAM-30m 1.3\,mm map is $\sim$\,650\,mJy (aperture 60$''$; Andr{\'e} 
et al. 1999), indicating that more than 95\% of the flux from the large-scale envelope around the two sources 
was resolved out by the SMA observations.
Furthermore, it must be noted that in the IRAM Plateau de Bure Interferometer (PdBI) 1.3\,mm continuum 
observations, only a single object was detected in IRAM\,04191, which is spatially coincident with source 
IRAM\,04191\,IRS (see Maury et al. 2010). It is likely that the PdBI observations missed the relatively faint 
source MMS (with about half of the IRS flux), due to the low flux ratio sensitivity in the IRAM\,04191 
observations (only about 0.6--0.9; Maury et al. 2010).

Assuming that the 1.3\,mm dust continuum emission is optically thin, the total gas mass ($M_{\rm gas}$) of 
the two sources is derived from their fluxes using the same method as described in Launhardt \& Henning
(1997). In the calculations, we adopt a dust opacity of $\kappa_{\rm d} = 0.5\,{\rm cm}^2\,{\rm g}^{-1}$ 
(Ossenkopf \& Henning 1994), a typical value for dense and cold molecular cloud cores. A dust temperature of 
$\sim$\,20\,K was adopted for IRAM\,04191\,MMS and IRS, which is derived from fitting the SEDs (see below). 
The relative uncertainties of the derived masses due to the calibration errors of the fluxes are $\pm$20\%. The 
total gas masses of IRAM\,04191\,IRS and MMS derived from the SMA maps are $\sim$\,0.011\,$M_\odot$ 
and $\sim$\,0.005\,$M_\odot$, respectively.

\section{DISCUSSION}

\subsection{The Spectral Energy Distributions}

In order to understand the properties of sources IRAM\,04191\,IRS and MMS, it is essential to construct 
their spectral energy distributions (SEDs), and then to derive their luminosities and temperatures. However, 
previous single-dish (sub-)\,millimeter continuum observations lack the resolution to resolve the two sources. 
Therefore, we simply assume a flux ratio of 2:1 at these wavelengths (from 160\,$\mu$m to 1.3\,mm), which is 
inferred from the SMA flux ratio of the two sources. Furthermore, due to the fact that no compact infrared 
emission was detected from MMS in the $Spitzer$ 3.6 to 70\,$\mu$m bands but infrared emission was 
detected at the position of IRS, we assume that all the infrared flux at wavelengths $\leq$\,90\,$\mu$m 
comes from source IRS. Given these assumptions, we show in Figure~2 the SEDs of sources IRS and MMS 
(see Dunham et al. 2008 for individual flux values). To derive luminosities and temperatures, we first interpolated 
and then integrated the SEDs, assuming spherical symmetry. Interpolation between the flux densities was done 
by a $\chi$$^2$ single-temperature grey-body fit to all points at $\lambda$\,$>$\,90\,$\mu$m, using the 
method described in Chen et al. (2008). A simple logarithmic interpolation was performed between all points 
at $\lambda$\,$\leq$\,90\,$\mu$m (including 3\,$\sigma$ upper limits for source MMS). From the SED fitting, 
we estimate $L_{\rm bol}$\,$\sim$\,0.08\,$L_\odot$ and $T_{\rm bol}$\,$\sim$\,25\,K for source IRS, and 
$L_{\rm bol}$\,$<$\,0.03\,$L_\odot$ and $T_{\rm bol}$\,$<$\,20\,K for source MMS. The dust temperature 
estimated for the two sources is 20$\pm$3\,K. Further observations, such as high angular resolution and high 
sensitivity continuum observations at wavelengths from far-infrared to (sub-)\,millimeter, are needed to constrain 
the SEDs of IRAM\,04191\,IRS and MMS, in order to derive more precisely their luminosities and temperatures.

\subsection{A Case of Rotational Fragmentation}

It has been suggested that the fragmentation of rotationally flattened cloud cores with initially flat density profiles, 
immediately after a phase of free-fall collapse, is one of the most efficient mechanisms for binary star formation 
(see reviews by Tohline 2002 and Goodwin et al. 2007). Numerical simulations find that rotating cloud cores
fragment if $\alpha$$_{\rm therm}$$\beta_{\rm rot}$ $\leq$ 0.12-0.15 (e.g., Tohline 1981), where 
$\alpha_{\rm therm}$ = $E_{\rm therm}$/$E_{\rm grav}$ is the initial thermal virial ratio ($E_{\rm therm}$ is the 
thermal kinetic and $E_{\rm grav}$ is the gravitational potential energy) and $\beta_{\rm rot}$ = $E_{\rm rot }$/$E_{\rm grav}$ 
is the initial rotational virial ratio ($E_{\rm rot}$ is the rotational kinetic energy). In particular, the simulations find 
that the level of rotation energy, i.e., $\beta_{\rm rot}$, plays an important role in this fragmentation process. For 
example, Boss (1999) found that rotating cloud cores fragment only when $\beta_{\rm rot}$ $>$ 0.01 (see also 
Machida et al. 2005).

Figure~3 shows the velocity field of IRAM\,04191, using the combined IRAM-30m and PdBI N$_2$H$^+$\,(1--0)
data from Belloche \& Andr{\'e} (2004) and derived with the same method used by Chen et al. (2007). The mean 
velocity map shows a clear continuous velocity gradient across the dense core, increasing from northwest to southeast, 
i.e., roughly along the line connecting sources MMS and IRS (see Figure~3a). As discussed in Chen et al. (2007), 
systematic velocity gradients found in protostellar cores/envelopes are usually dominated by either rotation or outflow. 
As seen in Figure~3a, the large angle ($>$\,70$^\circ$) between the gradient and the axis of the IRS outflow suggests 
that the observed velocity gradient in N$_{2}$H$^{+}$ is due to core rotation rather than outflow.
Figure~3b shows the spatial distribution of the N$_{2}$H$^{+}$ line widths in the map. The line widths are roughly 
constant within the core, and relatively large line widths are mainly seen in the central region of the core (roughly 
along the outflow axis), which might be caused by the outflow (see Chen et al. 2007 for more discussions).

From the velocity field derived from the N$_2$H$^{+}$\,(1--0) observations, we can estimate the ratios of rotational
and thermal energy to gravitational potential energy, using equations obtained from Chen et al. (2007; 2008):
\begin{equation}
\beta_{\rm rot} = 0.263 \times (\frac{R}{\rm pc})^2 \times (\frac{g}{\rm km\,s^{-1}\,pc^{-1}})^2 \times (\frac{\rm km\,s^{-1}}{\Delta v_{\rm obs}})^2,
\end{equation}
\begin{equation}
\alpha_{\rm therm} = 0.01 \times (\frac{T}{\rm K}) \times (\frac{\rm km\,s^{-1}}{\Delta v_{\rm obs}})^2,
\end{equation}
where $R$ is the core radius (the FWHM radius $\sim$\,20$''$ or 2800\,AU), $g$ is the velocity gradient 
($\sim$\,17\,km\,s$^{-1}$\,pc$^{-1}$), $T$ is the kinetic gas temperature (6--10\,K; see Belloche et al. 
2002), and $\Delta$$v_{\rm obs}$ is the observed mean line width (0.50--0.55\,km\,s$^{-1}$; see Figure~3b).
The estimated $\beta_{\rm rot}$ and $\alpha_{\rm therm}$ values for IRAM\,04191 are 0.06\,$\pm$0.02
and 0.3\,$\pm$\,0.1, respectively. Correcting for inclination ($i$\,=\,50$^\circ$; see Belloche et al. 2002), 
the estimated $\beta_{\rm rot}$ is then about 0.17, consistent with the value of $\sim$\,0.1--0.2 derived 
by Belloche et al. (2002).
It is clear from these values that IRAM\,04191 meets the classical fragmentation criterion of $\alpha$$_{\rm therm}$$\beta_{\rm rot}$ $\leq$ 
0.12, and the estimated $\beta_{\rm rot}$ value is far beyond the critical value suggested by numerical simulations. 
We thus suggest that the two sources in IRAM\,04191 were formed through the rotational fragmentation 
of an elongated dense core.
Furthermore, because of its clear rotational picture, close distance, and youth, as well as its other 
well-characterized properties derived from previous observations, IRAM\,04191 represents so far 
one of the best examples to observationally study the rotational fragmentation of dense cores.

Interestingly, Tohline (1981) pointed out that the semi-major axis of a binary formed through rotational 
fragmentation is

\begin{equation}
a_{\rm bin} = (\frac{4}{\pi}) \beta_{\rm rot} R_{\rm core}.
\end{equation}

\noindent For IRAM\,04191, its core radius derived from the IRAM-30m 1.3\,mm observations is about 
4200\,AU or 30$''$ (see Andr{\'e} et al. 1999). Adopting $\beta_{\rm rot}$ = 0.17, the predicted binary 
separation is $\sim$\,6.5$''$, close to the value measured in the SMA 1.3\,mm continuum images ($\sim$\,7.8$''$).

\subsection{The Role of Rotation in Binary Star Formation}

At this stage, it is instructive to compare IRAM\,04191 to other protostars with high angular resolution 
kinematics data, and then to observationally investigate the role of rotation in binary star formation. Based 
on our previous observations (Chen et al. 2007; 2008; 2009), we have collected high angular resolution 
N$_2$H$^+$\,(1--0) data for twelve Class\,0 protostars. Among those protostars, two protobinary systems, 
CB\,230 (Chen et al. 2007) and SVS\,13B (Chen et  al. 2009), show a similar rotational picture as seen in 
IRAM\,04191, and may represent another two cases of rotational fragmentation.
In addition, VLA NH$_3$ observations suggest core/envelope rotation in the protostars HH\,212-MMS 
(Wiseman et al. 2001) and HH\,211-mm (Tanner \& Arce 2011), both of which were also resolved into binary 
systems in high angular resolution observations (Codella et al. 2007; Lee et al. 2009). More recently, Tobin 
et al. (2011) present high angular resolution N$_2$H$^+$ and NH$_3$ data for another eleven protostars. 
With these kinematic data, we estimate the $\beta_{\rm rot}$ values for the individual sources (using the 
same method given above), and show in Figure~4 the distribution of $\beta_{\rm rot}$ versus core FWHM 
radius for those sources (more analyses based on these kinematics data will be presented in a forthcoming 
paper).

As seen in Figure~4, we find that most cores with binary systems formed therein have $\beta_{\rm rot}$ $>$ 1\%. 
In contrast, those cores with $\beta_{\rm rot}$ $<$ 1\%, such as the well-studied protostar L1157, remain as 
single protostars in the observations.
This is consistent with the theoretical simulations (see above) and indicates that the level of rotational energy in a 
dense core plays an important role in the fragmentation process.
Nevertheless, a few protostars with $\beta_{\rm rot}$ $>$ 1\% also remain as single (see Figure~4). We speculate
that these protostars are unresolved binary/multiple systems due to the limitation of angular resolution or sensitivity. 
Indeed, for example, our SMA HCO$^+$ and CO observations suggest that one such protostar, L1521F, is likely 
a multiple system with a quadrupolar outflow (X. Chen et al. in preparation). 
However, it must be noted that (1) the size of the sample in Figure~4 is still small and no statistically significant
conclusion can be drawn yet; and (2) more high angular resolution observations are needed to study a larger sample 
thoroughly and to investigate if other core properties, e.g., turbulence, density structure, and magnetic field, also 
have effects on the fragmentation process.

\section{SUMMARY}

We report the discovery of a binary system with an angular separation of 7.8\,$\pm$\,0.2$''$ in the well-known 
Class\,0 protostar IRAM\,04191, using the SMA observations. Complementary IRAM N$_2$H$^+$\,(1--0) data 
suggest that the binary system is formed through the rotational fragmentation of an elongated dense core.
Furthermore, comparisons between IRAM\,04191 and other protostellar cores suggest that most cores with a
binary system formed therein have ratios of rotational energy to gravitational energy $\beta_{\rm rot}$ $>$ 1\%, 
which is consistent with theoretical simulations and indicates that the level of rotational energy in a dense core 
plays an important role in the fragmentation process. 

\acknowledgments

We thank the anonymous referee for many insightful comments and suggestions. We thank A.~Belloche and P.~Andr{\'e} 
for providing their N$_2$H$^{+}$\,(1--0) data of IRAM\,04191+1522. This material is based on work supported by NSF 
grant AST-0845619 to HGA.

\clearpage

\clearpage


\begin{deluxetable}{lccccc}
\tabletypesize{\scriptsize} \tablecaption{\footnotesize SMA
1.3\,mm dust continuum results of IRAM\,04191\label{sma dust continuum}} \tablewidth{0pt} \tablehead{\colhead{Source}
&\colhead{R.A. \& Dec. (J2000)}  & \colhead{Flux$^a$} &\colhead{FWHM Sizes$^a$} & \colhead{$M_{\rm gas}$$^b$}\\
\cline{4-4}\colhead{Name}   & \colhead{[h\,:\,m\,:\,s,
$^{\circ}:\,':\,''$]} & \colhead{[mJy]}& \colhead{Maj.\,$\times$\,Min.} & \colhead{[$M_\odot$]}} \startdata

IRAM\,04191\,IRS     & 04:21:56.90, 15:29:46.4  & 9.3$\pm$2.0  & 3.6$''$\,$\times$\,3.5$''$ & 0.011$\pm$0.002 \\

IRAM\,04191\,MMS   & 04:21:56.37, 15:29:48.9 & 4.7$\pm$1.0 & 2.5$''$\,$\times$\,1.6$''$ & 0.005$\pm$0.001 \\

\enddata
\tablenotetext{a}{Flux and FWHM sizes derived from the Gaussian fitting.}
\tablenotetext{b}{Total gas mass; See text for dust temperature and opacity used.}
\end{deluxetable}

\clearpage

\begin{figure*}
\begin{center}
\includegraphics[width=16cm,angle=0]{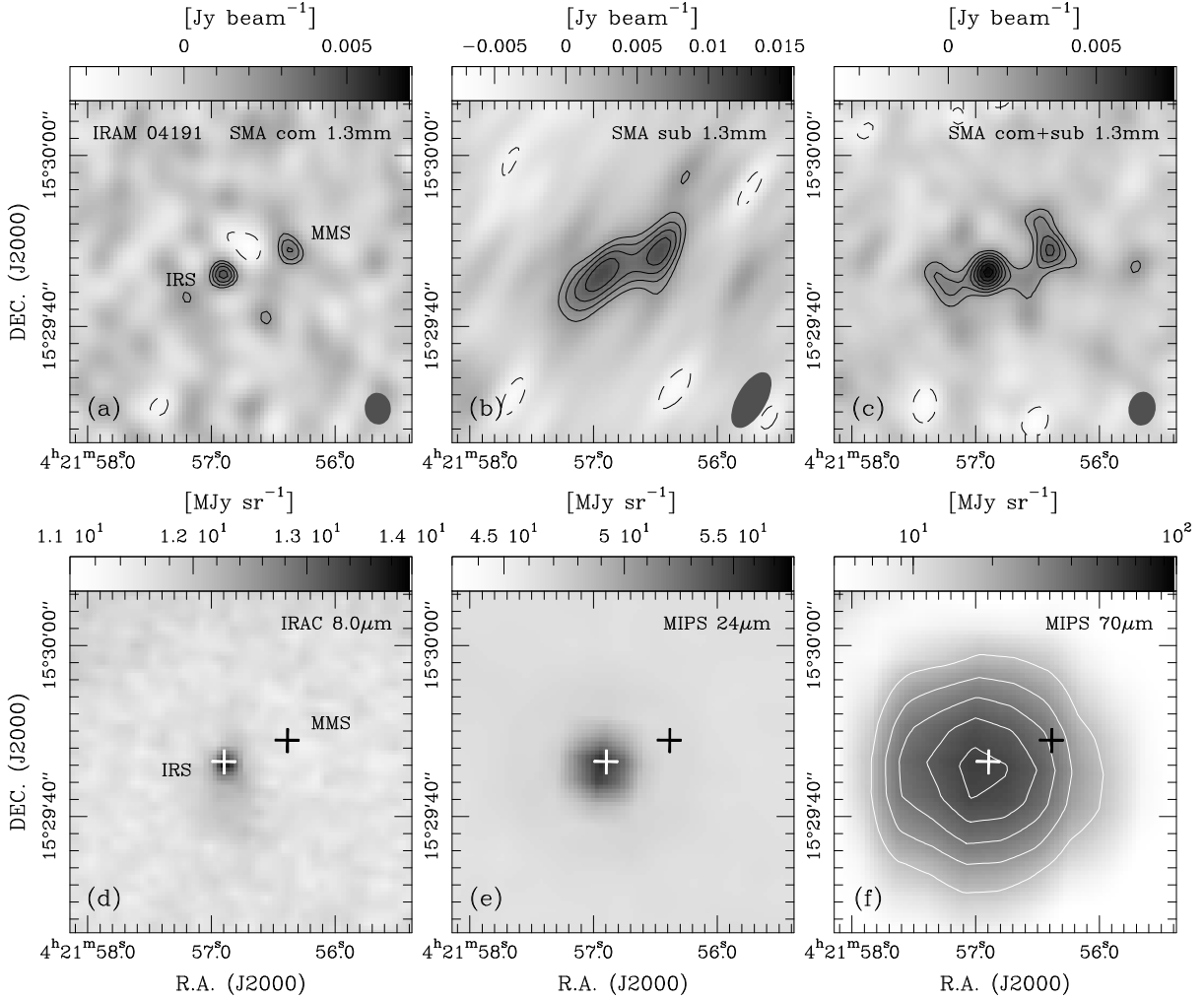}
\caption{(a) SMA 1.3\,mm dust continuum image of IRAM\,04191, taken with the SMA compact configuration.
The contours start at $\pm$3\,$\sigma$, and then increase from +3\,$\sigma$ in steps of $\sim$\,1\,$\sigma$ 
(1\,$\sigma$\,$\sim$\,0.70\,mJy\,beam$^{-1}$). The synthesized SMA beam is shown as a grey oval in the 
bottom right corner. (b) Same as Figure~a, but taken with the SMA subcompact configuration (1\,$\sigma$\,$\sim$\,1.6\,mJy\,beam$^{-1}$). 
(c) The SMA 1.3\,mm continuum image, using the combined compact and subcompact data (1\,$\sigma$\,$\sim$\,0.77\,mJy\,beam$^{-1}$).
(d) The $Spitzer$ IRAC 8.0\,$\mu$m image of IRAM\,04191. The two crosses show the positions of the two SMA
continuum sources IRS (white) and MMS (black). (e) Same as Figure~d, but the MIPS 24\,$\mu$m image. (f) Same
as Figure~d, but the MIPS 70\,$\mu$m image (the contours start at 16\,MJy\,sr$^{-1}$, and increase in steps of 10\,MJy\,sr$^{-1}$). 
The $Spitzer$ data of IRAM\,04191 are adopted from Dunham et al. (2006).\label{iram04191_sma_spitzer}}
\end{center}
\end{figure*}


\begin{figure*}
\begin{center}
\includegraphics[width=12cm, angle=0]{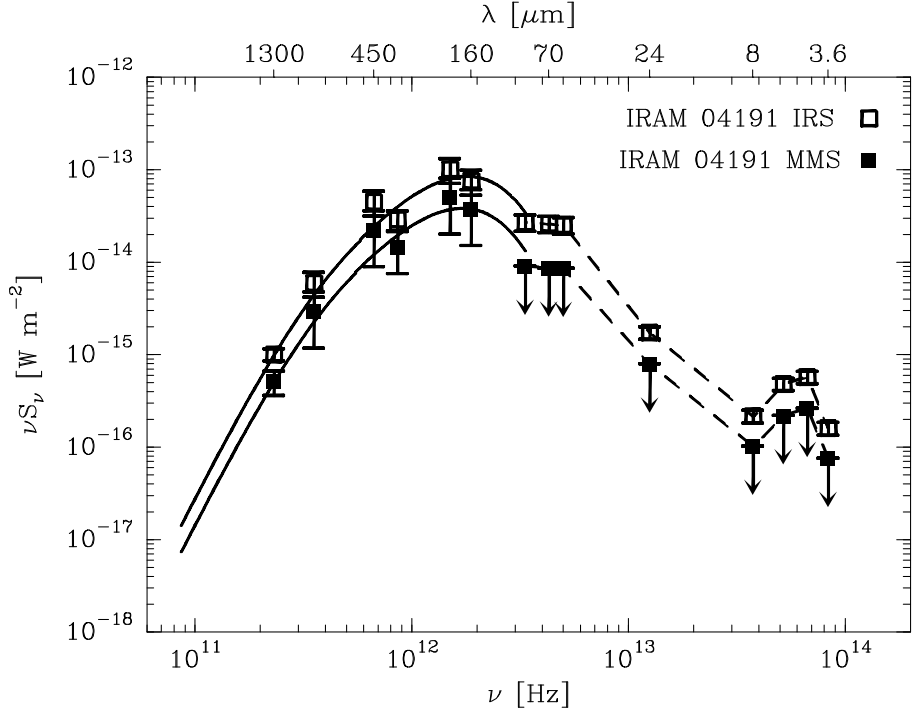}
\caption{Spectral energy distributions of IRAM\,04191 IRS and MMS. Solid lines show the best-fit for all points at 
$\lambda$ $>$ 90\,$\mu$m using a grey-body model $S_{\nu}$\,=\,$B_{\nu}$($T_{\rm d}$)(1\,$-$\,$e$$^{-\tau_\nu}$)$\Omega$, 
where $B_{\nu}$($T_{\rm d}$) is the Planck function at frequency $\nu$ and dust temperature $T_{\rm d}$, $\tau_\nu$ is the dust 
optical depth as a function of frequency $\tau$\,$\propto$\,$\nu$$^{1.8}$, and $\Omega$ is the solid angle of the source.
Dashed lines show a simple logarithmic interpolation performed between all points at $\lambda$ $\leq$ 90\,$\mu$m for 
the two sources (the datapoints of MMS at wavelengths $\leq$ 90\,$\mu$m are 3\,$\sigma$ upper limits derived from the
infrared images).\label{iram04191_sed}}
\end{center}
\end{figure*}

\begin{figure*}
\begin{center}
\includegraphics[width=11.5cm, angle=0]{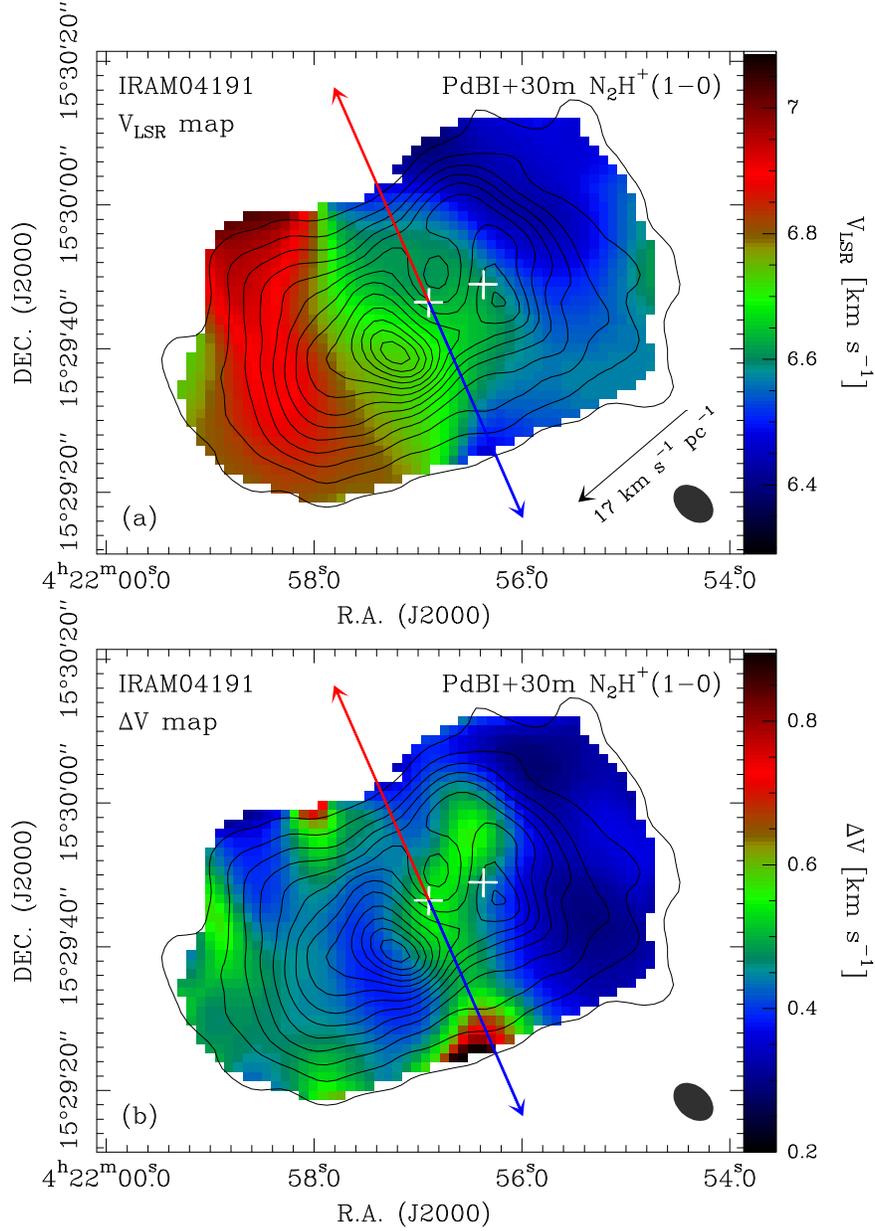}
\caption{N$_2$H$^+$\,(1--0) velocity fields of IRAM\,04191+1522, derived from the N$_2$H$^+$ data 
combining the IRAM-30m and PdBI observations (data from Belloche \& Andr{\'e} 2004). The top map
shows the mean velocity field (color shades), while the bottom map shows the spatial distribution of
line widths (color shades). In both maps, black contours show the integrated intensity, from 0.16 to 2.1 
in steps of 0.16 Jy\,beam$^{-1}$\,km\,s$^{-1}$ ($\sim$\,6 times the rms noise); white crosses show the 
positions of the SMA continuum sources IRS and MMS; red and blue arrows show the directions of the
red- and blue-shifted CO outflow lobes. The synthesized cleaned beam in the N$_2$H$^+$ observations 
is shown in the bottom right corner. In the mean velocity map, the black arrow in the bottom right shows 
the direction of the velocity gradient (P.A. $\sim$\,133$^\circ$).\label{velocity_field}}
\end{center}
\end{figure*}

\begin{figure*}
\begin{center}
\includegraphics[width=13cm, angle=0]{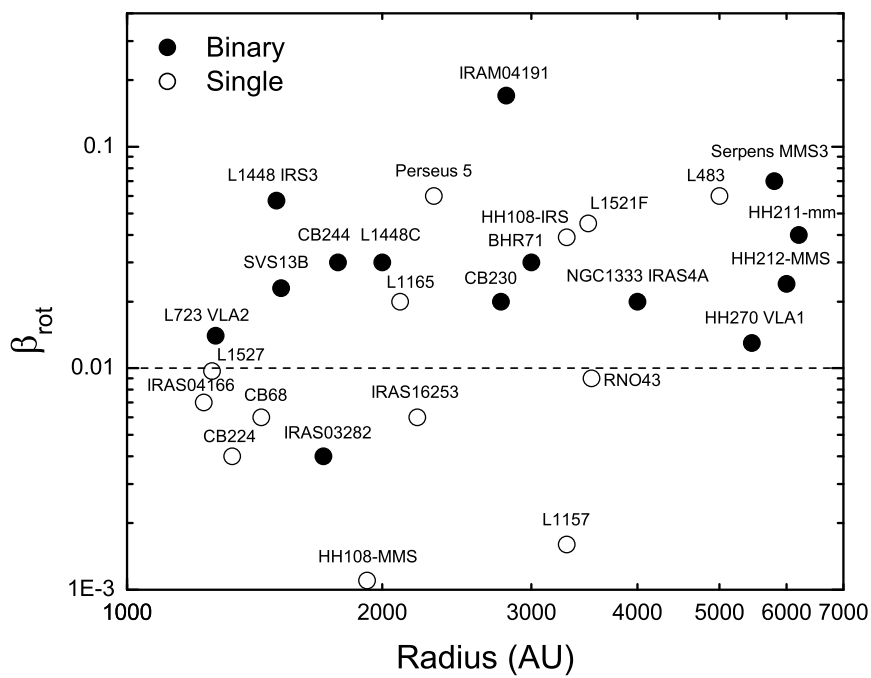}
\caption{Ratio of rotational energy to gravitational energy $\beta_{\rm rot}$ vs. core FWHM radius.
Data for the sources shown are adopted from Chen et al. (2007; 2008; 2009) and Tobin et al. (2011), 
except IRAM\,04191 (this work), NGC\,1333 IRAS\,4A (Belloche et al. 2006), HH\,212-MMS (Wiseman 
et al. 2001), HH\,211-mms (Tanner \& Arce 2011), and L1448C/L1448\,IRS3 (X.~Chen et al. in preparation). 
Note that except for IRAM\,04191, the values of $\beta_{\rm rot}$ in this plot are lower limits, as they are
not corrected for inclination.\label{beta_rot}}
\end{center}
\end{figure*}

\end{document}